\RequirePackage{amsmath}
\documentclass{iopart}

\usepackage{graphicx}
\usepackage{dcolumn}
\usepackage{bm}
\usepackage{amsmath}

\begin{document}

\title{Macroscopic fluxes and local reciprocal relation in second-order stochastic processes far from equilibrium}

\author{Hao Ge}

\address{Beijing International Center for Mathematical Research (BICMR) and Biodynamic Optical Imaging Center (BIOPIC), Peking University, Beijing, 100871, PRC.}

\address{E-mail: \mailto{haoge@pku.edu.cn}}

\begin{abstract}
Stochastic process is an essential tool for the investigation of the physical and life sciences at nanoscale. In the first-order stochastic processes widely used in chemistry and biology, only the flux of mass rather than that of heat can be well defined. Here we investigate the two macroscopic fluxes in second-order stochastic processes driven by position-dependent forces and temperature gradient. We prove that the thermodynamic equilibrium defined through the vanishing of macroscopic fluxes is equivalent to that defined via time reversibility at mesoscopic scale. In the small noise limit, we find that the entropy production rate, which has previously been defined by the mesoscopic irreversible fluxes on the phase space, matches the classic macroscopic expression as the sum of the products of macroscopic fluxes and their associated thermodynamic forces. Further we show that the two pairs of forces and fluxes in such a limit follow a linear phenomenonical relation and the associated scalar coefficients always satisfy the reciprocal relation for both transient and steady states. The scalar coefficient is proportional to the square of local temperature divided by the local frictional coefficient and originated from the second moment of velocity distribution along each dimension. This result suggests the very close connection between Soret effect (thermal diffusion) and Dufour effect at nano scale even far from equilibrium.

\end{abstract}


\maketitle

\tableofcontents

\section{Introduction}

Second-order stochastic processes describe the movement of macromolecules at nanoscale, in which the heat reservoir is coarse grained as two additional force terms, i.e. a frictional force and a random fluctuating force, into the deterministic Newtonian mechanics. When the frictional force dominates, the dynamics can be approximated by a first-order stochastic process, as always used for modeling the chemical and cellular dynamics. However, the real physics is in second order rather than in first order. That is why recently the nonequilibrium thermodynamics of second-order stochastic processes has caused so much interests \cite{Qian_Kim,Ford,hidden_epr,Ge_PRE2014,Polettini}.

We notice that most results of previous works in this area are interpreted in terms of mesoscopic fluxes and associated thermodynamic forces on the phase space rather than the macroscopic fluxes that are directly related to nonequilibrium thermodynamics, hence there are many fundamental properties which still remain to be answered. Hence in the present paper, we first address the question whether different definitions of thermodynamic equilibrium at mesoscopic and macroscopic scales are equivalent or not. Mesoscopic definition of thermodynamic equilibrium arises from time-reversibility of the stochastic process \cite{Ge_PRE2014}, which is equivalent to the  vanishing of irreversible velocity fluxes on the phase space, while the macroscopic definition of thermodynamic equilibrium is simply the vanishing of all macroscopic fluxes driven by various kinds of thermodynamic forces. Of course, it is trivial that the time reversibility implies the vanishing of macroscopic fluxes of mass and heat. However, the opposite direction is far from obvious because in going from mesoscopic scale to the macroscopic scale the irreversibility on the phase space might be erased when taking macroscopic averages. In the present paper, we will prove that the opposite direction still holds.

We then prove that in the small noise limit, the local entropy production rate defined through the mesoscopic irreversible velocity fluxes on the phase space can be expressed as the sum of the products of macroscopic fluxes and their associated thermodynamic forces. We further show the two pairs of forces and fluxes in such a limit satisfy a linear phenomenonical relation and the associated reciprocal relation always holds for both transient and steady states, which indicates the very close connection between Soret effect (thermal diffusion) and Dufour effect at nano scale even far from equilibrium.

\section{Model and previous results}

The motion of a Brownian particle at nanoscale driven by the position-dependent forces as well as spatial temperature gradient can be modeled by the $n$-dimensional second-order stochastic differential equation
\begin{eqnarray}
\frac{d\bm{X}}{dt}&=&\bm{V};\nonumber\\
m\frac{d\bm{V}}{dt}&=&-\eta(\bm{X})\bm{V}+G(\bm{X})+\xi(t)\label{Langevin_eq}
\end{eqnarray}
on the phase space, where $\xi(t)$ represents Gaussian white noise with position-dependent intensity
$D(\bm{x})$ and $m$ is the mass of the particle. For such a system, we can define an effective temperature $T(\bm{x})$ from the local Einstein relation, i.e. $D(\bm{x})=2\eta(\bm{x})k_BT(\bm{x})$. Here we restrict ourselves to the case that the effective temperature at each instantaneous position is imposed by the huge heat reservoir, which is independent of the fluctuation of the system (\ref{Langevin_eq}) \cite{VanKampen,hidden_epr}.

The corresponding Fokker-Planck equation for the evolution of probability distribution $\rho_t(\bm{x},\bm{v})=\mathbf{P}(\bm{X}(t)=\bm{x},\bm{V}(t)=\bm{v})$ in the phase space is \cite{Ge_PRE2014,Risken,Gardiner85}
\begin{equation}
\frac{\partial}{\partial t}\rho_t=-\nabla_{\bm{x}}\cdot \bm{j}_{\bm{x}}-\nabla_{\bm{v}}\cdot \bm{j}_{\bm{v}},\label{FP1}
\end{equation}
where the coordinate flux $\bm{j}_{\bm{x}}=\bm{v}\rho_t$ and the velocity flux $\bm{j}_{\bm{v}}=\frac{-\eta(\bm{x})\bm{v}+G(\bm{x})}{m}\rho_t-\frac{D(\bm{x})}{2m^2}\nabla_{\bm{v}}\rho_t$.


It is known that the velocity flux $\bm{j}_{\bm{v}}$ can be decomposed into an irreversible term $\bm{j}^{ir}_{\bm{v}}=\left(-\frac{\eta(\bm{x})}{m}\bm{v}-\frac{D(\bm{x})}{2m^2}\nabla_{\bm{v}}\log\rho_t\right)\rho_t$ associated with the thermodynamic force $\bm{f}^{ir}_{\bm{v}}=\frac{\bm{j}^{ir}_{\bm{v}}}{\rho_t}$ and a reversible term $\bm{j}^{rev}_{\bm{v}}=\frac{G(\bm{x})}{m}\rho_t$ \cite{Ford,Risken,Ge_PRE2014}. The entropy production rate, recently defined, is just the average of the conditional second moment of $\bm{f}^{ir}_{\bm{v}}$ at given position \cite{Qian_Kim,Ford,Ge_PRE2014}, i.e.
\begin{eqnarray}
epr&=k_B\int\frac{2m^2}{D(\bm{x})}\langle \left(\bm{f}^{ir}_{\bm{v}}\right)^2\rangle_{\bm{x}}\hat{\rho}_t(\bm{x})d\bm{x},\label{epr}
\end{eqnarray}
where
$\langle \left(\bm{f}^{ir}_{\bm{v}}\right)^2\rangle_{\bm{x}}=\int \left(\bm{f}^{ir}_{\bm{v}}\right)^2\frac{\rho_t}{\hat{\rho}_t(\bm{x})} d\bm{v}$
is the second moment of $\bm{f}^{ir}_{\bm{v}}$ given $\bm{x}$,
and $\hat{\rho}_t(\bm{x})=\int \rho_t d\bm{v}$ is the marginal distribution at the spatial coordinate.

In a previous work, we have already known that the time reversibility for such a second-order stochastic system is equivalent to thermodynamic equilibrium, which means there is no nonconservative force (mechanical equilibrium) and no temperature gradient (thermal equilibrium), i.e. the position-dependent force $G(\bm{x})$ associates with a potential $U(\bm{x})$ and the temperature profile $T(\bm{x})$ is independent of $\bm{x}$ \cite{Ge_PRE2014}. Provided it is at steady state, the entropy production rate vanishes if and only if the stochastic process is at thermodynamic equilibrium \cite{Ford,Ge_PRE2014}.


Regarding the first law of thermodynamics along the spatial coordinate, the evolution of the spatial density of kinetic energy $E_t^{kinetic}(\bm{x})=\int \frac{1}{2}m\bm{v}^2\rho_t(\bm{x},\bm{v})d\bm{v}$ is \cite{Ge_PRE2014}
\begin{equation}
\frac{d}{dt}E_t^{kinetic}(\bm{x})+\nabla_{\bm{x}}\cdot\bm{J}_{\bm{x}}^{kinetic}=W(\bm{x},t)-Q(\bm{x},t),\label{flow_kinetic}
\end{equation}
in which $\bm{J}_{\bm{x}}^{kinetic}=\int \frac{1}{2}m\bm{v}^2\bm{j}_{\bm{x}}d\bm{v}$ is the spatial flux of kinetic energy, $Q(\bm{x},t)=-\int \bm{f}^{ir}_{\bm{v}}\bm{v}\rho_td\bm{v}$ is the spatial heat dissipation density, and $W(\bm{x},t)=G(\bm{x})\bm{J}_{\bm{x}}$ is the spatial density of work done upon the system at position $\bm{x}$. $\bm{J}_{\bm{x}}$ is the integrated spatial fluxes $\bm{J}_{\bm{x}}=\int \bm{j}_{\bm{x}}d\bm{v}$.




\section{Thermodynamic equilibrium in terms of vanishing macroscopic fluxes}

At thermodynamic equilibrium, the second-order dynamics is called Klein-Kramers equation, taking Maxwell-Boltzmann distribution as its steady-state distribution \cite{Kramers}. Hence in this case, it is quite trivial that the entropy production rate $epr$, the spatial density of heat dissipation $Q(\bm{x})$, the spatial flux of kinetic energy $\bm{J}_{\bm{x}}^{kinetic}$ and the integrated spatial fluxes $\bm{J}_{\bm{x}}$ all vanish.

Moreover, we can define the heat flux at each spatial coordinate $\bm{x}$ following Groot and Mazur \cite{Groot_Mazur1984} as
\begin{equation}
\bm{J}_q(\bm{x})=\bm{J}_{\bm{x}}^{kinetic}-E_t^{kinetic}(\bm{x})\bar{\bm{v}}_{\bm{x}},\nonumber
\end{equation}
in which the averaged velocity at spatial coordinate $\bm{x}$ is
$\bar{\bm{v}}_{\bm{x}}=\frac{\int \bm{v}\rho_td\bm{v}}{\hat{\rho}_t(\bm{x})}=\frac{\bm{J}_x}{\hat{\rho}_t(\bm{x})}$.

Hence at steady state, if both the macroscopic fluxes of heat and mass vanish, i.e. $\bm{J}_{q}(\bm{x})=\bm{J}_{\bm{x}}=0$, then $\bm{J}^{kinetic}_{\bm{x}}$ also vanish. According to the definition of the local densities of work $W^{ss}(\bm{x})$ and the evolution of kinetic energy along the spatial coordinate (\ref{flow_kinetic}), we know that the local densities of heat dissipation $Q^{ss}(\bm{x})$ vanish at each spatial coordinate, which is followed by the vanishing of entropy production rate due to the fact that $epr=\int \frac{Q^{ss}(\bm{x})}{T(\bm{x})}d\bm{x}$ at steady state \cite{Ge_PRE2014}.

Therefore, both the macroscopic fluxes of heat and mass vanish at each spatial coordinate if and only if the system is at thermodynamic equilibrium. This result confirms the equivalent definitions of thermodynamic equilibrium at the mesoscopic and macroscopic scales. The top-down proof from the macroscopic definition to the mesoscopic definition is not that obvious, since it is not simply the reversed proof of the opposite direction.

In addition, we notice that the local densities of heat dissipation and kinetic energy are related, i.e.
\begin{equation}
Q(\bm{x},t)=\frac{2\eta(\bm{x})}{m}\left[E_t^{kinetic}(\bm{x})-\frac{n}{2}k_BT(\bm{x})\hat{\rho}_t(\bm{x})\right],\nonumber
\end{equation}
hence at steady state, thermodynamic equilibrium is equivalent to the local equipartition theorem, i.e. $E^{kinetic}(\bm{x})=\frac{n}{2}k_BT(\bm{x})\hat{\rho}(\bm{x})$, in which $n$ is the dimension. It implies that in the case of the second-order stochastic process (\ref{Langevin_eq}), once the local equipartition theorem is shown to be valid at each spatial coordinate, the temperature profile in fact is spatially uniform and the whole system is at thermodynamic equilibrium.

\section{Linear relation between pairs of macroscopic fluxes and forces}

\subsection{Leading order of $\eta$ in the small noise limit}

In the limit of small inertia,  the marginal distribution of the spatial coordinate $\hat{\rho}_t(\bm{x})$ satisfies the corresponding Smoluchowski equation \cite{Smoluchowski,Gardiner85,hidden_epr,anti-Ito}:
\begin{equation}
\frac{\partial \hat{\rho}_t(\bm{x})}{\partial t}=-\nabla_{\bm{x}}\cdot\bm{J}_{\bm{x}}^{over},\label{Fokker-Planck-Smol}
\end{equation}
in which the overdamped spatial flux of mass $\bm{J}_{\bm{x}}^{over}(\bm{x})=\frac{G(\bm{x})}{\eta(\bm{x})} \hat{\rho}_t(\bm{x})-\frac{1}{\eta (\bm{x})}\nabla_{\bm{x}}\left[k_BT(\bm{x})\hat{\rho}_t(\bm{x})\right]$.

The leading order of $\rho_t(\bm{x},\bm{v})$ with respect to $\eta$ is \cite{Ge_PRE2014}
\begin{eqnarray}
&&\rho_t(\bm{x},\bm{v})\nonumber\\
&=&\hat{\rho}_t(\bm{x})w(\bm{v}|\bm{x})+w(\bm{v}|\bm{x})\frac{m\bm{v}}{k_BT(\bm{x})}\cdot\bm{J}_{\bm{x}}^{over}(\bm{x})\nonumber\\
&&+\hat{\rho}_t(\bm{x})w(\bm{v}|\bm{x})\frac{m\bm{v}\cdot\nabla_{\bm{x}}T(\bm{x})}{\eta(\bm{x})k_BT^2(\bm{x})}\left[\frac{n+2}{6}k_BT(\bm{x})-\frac{m\bm{v}^2}{6}\right],\nonumber\\
\label{app_dist}
\end{eqnarray}
in which $w(\bm{v}|\bm{x})$ as the locally approximated Maxwell-Boltzmann distribution $w(\bm{v}|\bm{x})=\frac{1}{\left(2\pi\frac{k_BT(\bm{x})}{m}\right)^{n/2}}e^{-\frac{m\bm{v}^2}{2k_BT(\bm{x})}}$.

From Eq. (\ref{app_dist}), we have got the leading order of local entropy production rate with respect to $\eta$ \cite{hidden_epr,Ge_PRE2014}
\begin{equation}
epr(\bm{x},t)\approx epr^{over}(\bm{x},t)+\Xi^{over}(\bm{x},t),\nonumber
\end{equation}
in which $epr^{over}(\bm{x},t)=\frac{\eta(\bm{x})}{T(\bm{x})}\left(\frac{\bm{J}^{over}_{\bm{x}}(\bm{x})}{\hat{\rho}_t(\bm{x})}\right)^2\hat{\rho}_t$ is just the local entropy production rate defined for the overdamped dynamics associated with (\ref{Fokker-Planck-Smol})\cite{JQQ}, and  $\Xi^{over}(\bm{x},t)=\frac{n+2}{6}k^2_B\frac{\left[\nabla_{\bm{x}}T(\bm{x})\right]^2}{\eta(\bm{x})T(\bm{x})}\hat{\rho}_t$ is regarded as the anomalous term\cite{hidden_epr,Ge_PRE2014}.

Also from Eq. (\ref{app_dist}), we can get the leading order of heat flux with respect to $\eta$
\begin{equation}
\bm{J}_q(\bm{x})\approx k_BT(\bm{x})\bm{J}_{\bm{x}}^{over}-\frac{n+2}{6}\frac{k_B^2T(\bm{x})}{\eta(\bm{x})}\left[\nabla_{\bm{x}}T(\bm{x})\right]\hat{\rho}_t(\bm{x}),\nonumber
\end{equation}
noticing that $\bm{J}_{\bm{x}}^{over}$ is at the order of $\frac{1}{\eta(\bm{x})}$.



\subsection{Reciprocal relation}

Denote $\bm{X}_q(\bm{x})=\nabla_{\bm{x}}\left(\frac{1}{T(\bm{x})}\right)$ as the local thermodynamic force associated with heat flux $\bm{J}_q(\bm{x})$, and $\bm{X}_p(\bm{x})$ as the local thermodynamic force for the flux of mass $\bm{J}_x^{over}(\bm{x})$. In order to keep the well-known macroscopic expression of entropy production rate as the sum of the products of pairs of macroscopic fluxes and related thermodynamic forces \cite{Prigogine68}, i.e.
\begin{equation}
epr(\bm{x},t)\approx \bm{X}_p(\bm{x})\cdot \bm{J}_x^{over}+\bm{X}_q(\bm{x})\cdot \bm{J}_q(\bm{x}),\label{phenomenonical}
\end{equation}
we can arrive at the expression of $\bm{X}_p(\bm{x})$, i.e.
\begin{equation}
\bm{X}_p(\bm{x})=\frac{G(\bm{x})-k_BT(\bm{x})\nabla_{\bm{x}}\log \hat{\rho}_t(\bm{x})}{T(\bm{x})}.\nonumber
\end{equation}

Consequently, we find that these pairs of fluxes and forces satisfy a linear relation
\begin{eqnarray}
\bm{J}_q(\bm{x})&=&\bm{L}_{qx}\cdot \bm{X}_p+\bm{L}_{qq}\cdot \bm{X}_q;\nonumber\\
\bm{J}_x^{over}&=&\bm{L}_{xx}\cdot \bm{X}_p+\bm{L}_{xq}\cdot \bm{X}_q,\label{reciprocal_relation}
\end{eqnarray}
in which the scalar coefficients $\bm{L}_{qx}=\bm{L}_{xq}=\frac{k_B\left(T(\bm{x})\right)^2}{\eta(\bm{x})}$, $\bm{L}_{qq}=\frac{n+8}{6}\frac{k_B^2\left(T(\bm{x})\right)^3}{\eta(\bm{x})}\hat{\rho}_t(\bm{x})$,
and $\bm{L}_{xx}=\frac{T(\bm{x})}{\eta(\bm{x})}\hat{\rho}_t(\bm{x})$. These coefficients are unique.

Eq. (\ref{reciprocal_relation}) implies the reciprocal relation between Soret effect(thermal diffusion) and Dufour effect can even be valid far from equilibrium. Soret effect is the phenomenon that the temperature gradient can cause a flux of mass and Dufour effect is the heat flux caused by concentration gradient respectively  \cite{Soret,Dufour,Duhr}. Noticing that when $G(\bm{x})=0$, the thermodynamic force $\bm{X}_p=-k_B\nabla_{\bm{x}}\log\hat{\rho}_t(\bm{x})$ is just the concentration gradient.

The derivation of reciprocal coefficients in near-equilibrium systems was first derived by Onsager and further generalized to systems including both even and odd variables by Casimir in terms of autocorrelation coefficients \cite{Onsager,Casimir}. The proof is generally based on the principle of microscopic reversibility \cite{Onsager,Casimir,Groot_Mazur1984}, which is only valid for systems slightly deviated from equilibrium. Recently, there are also several developments along this direction, which discovers interesting symmetry relations of fluctuation \cite{Gaspard2004}. However, the linear reciprocal relation of Onsager is still only valid in systems close to the local thermodynamic equilibrium, i.e. with rather small thermodynamic forces such as the temperature gradient.

For stochastic processes, time-reversibility is always equivalent to thermodynamic equilibrium \cite{JQQ,Ge_PRE2014}, which gives the foundation of Onsager's reciprocal relations. However, the converse is not necessary to be always true, i.e. probably there are certain pairs of fluxes and forces that always satisfy the reciprocal relation, even beyond equilibrium. Indeed, reciprocal relation has already been shown to be valid for certain class of irreversible interacting-particle systems, which should be carefully constructed \cite{Landim1996}. Here we show that in such a general second-order stochastic system (\ref{Langevin_eq}), the reciprocal relation naturally holds in general for the fluxes and thermodynamic forces of mass and heat, even in the far-from-equilibrium case.

Our linear relation (\ref{reciprocal_relation}) is quite local, which not only means the thermodynamic fluxes and forces considered here are all at a local spatial position, but also the scalar coefficients can only be invariant under small perturbation of the local temperature gradient $\nabla_{\bm{x}}T(\bm{x})$ or external force $G(\bm{x})$, while keeping the local temperature $T(\bm{x})$ and local transient density of mass $\hat{\rho}_t(\bm{x})$. Actually, the response coefficients in the original reciprocal relation derived by Onsager \cite{Onsager} are also only invariant under small perturbation of the system.

Next we would like to understand the mesoscopic origin of the phenomenonical coefficient $\bm{L}_{qx}=\bm{L}_{xq}$. The term $-\frac{\nabla_{\bm{x}}k_BT(\bm{x})}{\eta(\bm{x})}$ in $\bm{J}_{\bm{x}}^{over}$, where $\bm{L}_{xq}$ comes from, emerges in the derivative of the prefactor of $w(\bm{v}|\bm{x})$ with respect to $\bm{x}$ as we do the multiscale expansion for deriving the overdamped Smoluchowski equation (\ref{Fokker-Planck-Smol}) \cite{anti-Ito}. The prefactor of the Gaussian distribution $w(\bm{v}|\bm{x})$ is just proportional to the square root of the second moment of velocity along each dimension.

On the other hand, in the $n$ dimensional case,
the flux of kinetic energy can be decomposed as
\begin{eqnarray}
\bm{J}_{\bm{x}}^{kinetic}&=&\hat{\rho}_t(\bm{x})\langle\frac{1}{2}m\bm{v}^2\bm{v}\rangle_{\bm{x}}\nonumber\\
&=&\bm{J}_{\bm{1x}}^{kinetic}+\bm{J}_{\bm{2x}}^{kinetic}+\bm{J}_{\bm{3x}}^{kinetic}+\bm{J}_{\bm{4x}}^{kinetic},\nonumber
\end{eqnarray}
in which
$\bm{J}_{\bm{1x}}^{kinetic}=E_{\bm{x}}^{kinetic}\frac{\bm{J}_{\bm{x}}}{\hat{\rho}_t(\bm{x})}$, $\bm{J}_{\bm{2x}}^{kinetic}=\hat{\rho}_t(\bm{x})m\langle(\bm{v}\cdot\bar{\bm{v}})\bm{v}\rangle_{\bm{x}}$, $\bm{J}_{\bm{3x}}^{kinetic}=\hat{\rho}_t(\bm{x})\frac{1}{2}m\langle(\bm{v}-\bar{\bm{v}})^2(\bm{v}-\bar{\bm{v}})\rangle_{\bm{x}}$ and
$\bm{J}_{\bm{4x}}^{kinetic}=-\hat{\rho}_t(\bm{x})m\langle(\bar{\bm{v}})^2\bar{\bm{v}}\rangle_{\bm{x}}$.
Here $\langle\cdot\rangle_{\bm{x}}$ means the mean value under the conditional probability $\frac{\rho_t(\bm{x},\bm{v})}{\hat{\rho}_t(\bm{x})}$ with respect to $\bm{v}$ for fixed $\bm{x}$.

According to Eq. (\ref{app_dist}), in the small noise limit, we have $\bm{J}_{\bm{1x}}^{kinetic}\approx\frac{n}{2}k_BT(\bm{x})\bm{J}_{\bm{x}}^{over}$, $\bm{J}_{\bm{2x}}^{kinetic}\approx k_BT(\bm{x})\bm{J}_{\bm{x}}^{over}(\bm{x})$,
$\bm{J}_{\bm{3x}}^{kinetic}\approx\bm{J}_q(\bm{x})-k_BT(\bm{x})\bm{J}_{\bm{x}}^{over}(\bm{x})$ and
$\bm{J}_{\bm{4x}}^{kinetic}\approx 0$ upto the order $\frac{1}{\eta}$. It implies that the phenomenonical coefficient $\bm{L}_{qx}$
is just from $\bm{J}_{\bm{2x}}^{kinetic}$, in which the cross terms between different dimensions all vanish in the small noise limit and the remain terms are only the second moment of velocity along each dimension.

Meixner has shown that the reciprocal relations are invariant under certain transformations of the fluxes and thermodynamic forces \cite{Meixner}. In our case, the flux of mass $\bm{J}_x^{over}$ and thermodynamic force of heat $\bm{X}_q$ are defined conventionally and physically, which eliminates the possible self-contradictory of reciprocal relations if one can define new fluxes and forces from linear combination of the already defined fluxes and forces \cite{Coleman1960}.

There are several ways to choose alternative $\bm{J}_q(\bm{x})$ and $\bm{X}_p(\bm{x})$, but must keep the relation (\ref{phenomenonical}). For example, we can define the thermodynamic force of mass flux as $X^{'}_p(\bm{x})=\frac{\eta(\bm{x})\bm{J}_{\bm{x}}^{over}}{T(\bm{x})\hat{\rho}_t(\bm{x})}$ and the heat flux as
$\bm{J}_{q}^{'}(\bm{x})=\bm{J}_q(\bm{x})-k_BT(\bm{x})\bm{J}_{\bm{x}}^{over}$, in which the reciprocal relation still holds but with vanishing reciprocal coefficients.


In the case of conservative force, i.e. $G(\bm{x})=-\nabla_{\bm{x}}\phi(\bm{x})$, we can define a generalized chemical potential \cite{Rubi2005}
$$\mu(\bm{x})=\phi(\bm{x})+k_BT(\bm{x})\log \hat{\rho}_t(\bm{x}),$$
and we can clearly see that $\bm{X}_p$ is just the negative gradient of the chemical potential under constant temperature divided by the temperature, i.e. $\bm{X}_p(\bm{x})=-\frac{\left[\nabla_{\bm{x}}\mu(\bm{x})\right]_{T(\bm{x})}}{T(\bm{x})}$. It is just the standard definition of thermodynamic force conjugate to the heat flux $\bm{J}_q(\bm{x})$ \cite{Groot_Mazur1984}. This derivative does not depend on the standard-state contribution, i.e. invariant if we add a constant to the potential $\phi(\bm{x})$. Moreover, if we add the contribution of the potential energy $\phi(\bm{x})$ into both $\bm{J}_{\bm{x}}^{kinetic}$ and $E_t^{kinetic}(\bm{x})\bar{\bm{v}}_{\bm{x}}$, the expression of
heat flux $\bm{J}_q(\bm{x})$ that we used is invariant, hence it is also called the measurable heat flux \cite{Groot_Mazur1984}.

The total heat flux can be defined as the sum of the measurable heat flux and the flux of potential energy \cite{Groot_Mazur1984}, i.e.
$$\bm{J}_q^{''}=\bm{J}_q+\phi(\bm{x})\bm{J}_{\bm{x}}^{over}.$$
For diffusing substances, in fact the concept of heat flux can be defined in different ways, but leaves all physical results unchanged \cite{Groot_Mazur1984}.
In order to keep the relation (\ref{phenomenonical}), we find that the corresponding thermodynamic force of the mass flux is just
$\bm{X}^{''}_p=-\nabla_{\bm{x}}\frac{\mu(\bm{x})}{T(\bm{x})}$.

In this case, there is still a linear relation between the pairs of thermodynamic fluxes and forces
\begin{eqnarray}
\bm{J}_q^{''}(\bm{x})&=&\bm{L}^{''}_{qx}\cdot \bm{X}^{''}_p+\bm{L}^{''}_{qq}\bm{X}_q\nonumber\\
\bm{J}_x^{over}&=&\bm{L}^{''}_{xx}\cdot \bm{X}^{''}_p+\bm{L}^{''}_{xq}\cdot \bm{X}_q,
\end{eqnarray}
in which $\bm{L}^{''}_{xx}=\frac{T(\bm{x})}{\eta(\bm{x})}\hat{\rho}_t(\bm{x})$, $L^{''}_{qq}=\left[\frac{n+8}{6}\left(k_BT(\bm{x})\right)^2+2\phi(\bm{x})k_BT(\bm{x})+\phi^2(\bm{x})\right]\frac{T(\bm{x})}{\eta(\bm{x})}\hat{\rho}_t(\bm{x})$,
and the reciprocal coefficients $\bm{L}^{''}_{qx}=\bm{L}^{''}_{xq}=(k_BT(\bm{x})+\phi(\bm{x}))\frac{T(\bm{x})}{\eta(\bm{x})}\hat{\rho}_t(\bm{x})$.

The above coefficients depend on $\phi(\bm{x})$, which implies that both the total heat flux and these coefficients are not independent of the frame of references, i.e. not invariant if we add a constant number to $\phi(\bm{x})$. It suggests that choosing the measurable heat flux is more beneficial and thermodynamically robust.



\section{Summary and discussion}

The concept of entropy production rates in nonequilibrium thermodynamics dated back to the beginning of the 20th century, which is followed by the development of concrete expressions in various kinds of physical processes at the macroscopic scale and summarized as the phenomenonical relation (\ref{phenomenonical}) \cite{Prigogine68}. Recently, with the help of advanced experimental techniques, people can directly observe the stochastic processes at the nano scale \cite{Xie11}, hence the nonequilibrium thermodynamics of stochastic processes caught a lot of interests in physics and physical chemistry \cite{Stoc_thermo}. The entropy production rate at such nanoscale is typically defined from time reversibility, and expressed by mesoscopic fluxes and associated thermodynamic forces \cite{JQQ,Stoc_thermo,Seifert05}. Therefore, the linking between the two definitions at different scales remains to be verified and not obvious at all.

In the present letter, we have rigorously shown that the two definitions of thermodynamic equilibrium and entropy production rates are consistent in a general class of second-order stochastic processes. And in the overdamped limit, it is found that the macroscopic fluxes and associated thermodynamic forces satisfy a local linear relation with symmetric coefficients even far from equilibrium. This observation might indicate some previously unknown intrinsic coupling between the Soret and Dufour effects, beyond thermodynamic equilibrium.

In addition, as early as in Kramers' seminal work \cite{Kramers}, he already mentioned that the overdamped approximation is only valid if the external force $G(\bm{x})$ is almost constant on the thermal length scale $\sqrt{k_BT(\bm{x})/m}/\eta(\bm{x})$ \cite{Hanggi90}. Hence the linear relation between macroscopic fluxes and associated thermodynamic forces could possibly be violated in nano devices that do not satisfy this requirement.

\section{Acknowledgement}

We would like to thank
Dick Bedeaux, Matteo Polettini and Hong Qian for helpful discussion. H. Ge is supported by NSFC 10901040, 21373021 and the Foundation for the Author of National Excellent Doctoral Dissertation of China (No. 201119).

\end{document}